\def\papertitle{Sound texture synthesis using Convolutional Neural Networks}
\def\paperauthorA{Hugo Caracalla}
\def\paperauthorB{Axel Roebel}
\def\paperauthorC{Author Three}
\def\paperauthorD{Author Four}

\documentclass[twoside,a4paper]{article}
\usepackage{dafx_19}
\usepackage{amsmath,amssymb,amsfonts,amsthm}
\usepackage{euscript}
\usepackage[latin1]{inputenc}
\usepackage[T1]{fontenc}
\usepackage{ifpdf}

\usepackage[english]{babel}
\usepackage{caption}
\usepackage{subfig} 
\usepackage{color}

\ninept

\usepackage{times}

\newif\ifpdf
\ifx\pdfoutput\relax
\else
   \ifcase\pdfoutput
      \pdffalse
   \else
      \pdftrue
\fi

\ifpdf 
  \usepackage[pdftex,
    pdftitle={\papertitle},
    pdfauthor={\paperauthorA, \paperauthorB, \paperauthorC, \paperauthorD},
    colorlinks=false, 
    bookmarksnumbered, 
    pdfstartview=XYZ 
  ]{hyperref}
  \usepackage[pdftex]{graphicx}
\else 
  \usepackage[dvips]{epsfig,graphicx}
  \usepackage[dvips,
    colorlinks=false, 
    bookmarksnumbered, 
    pdfstartview=XYZ 
  ]{hyperref}
\fi
  \usepackage[hypcap=true]{caption}
\title{\papertitle}

\twoaffiliations{
\paperauthorA}
{UMR STMS 9912\\
Sorbonne Universit\'e, IRCAM, CNRS,\\
Paris, France\\
{\tt \href{mailto:hugo.caracalla@ircam.fr}{hugo.caracalla@ircam.fr}}
}
{\paperauthorB}
{UMR STMS 9912\\
IRCAM, Sorbonne Universit\'e, CNRS,\\
Paris, France\\
{\tt \href{mailto:axel.roebel@ircam.fr}{axel.roebel@ircam.fr}}
}

\def\hyph{-\penalty0\hskip0pt\relax}
\def\slash{/\penalty0\hskip0pt\relax}

\begin{document}
\ifpdf 
  \DeclareGraphicsExtensions{.png,.jpg,.pdf}
\else  
  \DeclareGraphicsExtensions{.eps}
\fi

\maketitle

\begin{abstract}
The following article introduces a new parametric synthesis algorithm for sound textures inspired by existing methods used for visual textures. Using a 2D Convolutional Neural Network (CNN), a sound signal is modified until the temporal cross-correlations of the feature maps of its log\hyph{}spectrogram resemble those of a target texture. We show that the resulting synthesized sound signal is both different from the original and of high quality, while being able to reproduce singular events appearing in the original. This process is performed in the time domain, discarding the harmful phase recovery step which usually concludes synthesis performed in the time-frequency domain. It is also straightforward and flexible, as it does not require any fine tuning between several losses when synthesizing diverse sound textures. A way of extending the synthesis in order to produce a sound of any length is also presented, after which synthesized spectrograms and sound signals are showcased. We also discuss on the choice of CNN, on border effects in our synthesized signals and on possible ways of modifying the algorithm in order to improve its current long computation time.
\end{abstract}

\section{Introduction}
\label{sec:intro}

The main difficulties encountered in sound texture synthesis become apparent when trying to properly define them. While examples of textures easily come to mind (e.g. environmental noises such as wind or rain, crowd hubbub, engine sounds, birds singing, etc.), pinpointing their common factors proves much harder: randomness seems to be one, along with a "background" aspect caused by an important number of indistinguishable small audio events happening at once. But this is not all there is to it, since we would still tend to call a sound including small occasional events happening in the foreground a texture. Hence, the definition offered by Saint-Arnaud \cite{saint1995classification}, summed up by Schwartz \cite{schwarz2011}, of "a superposition of small audio atoms overlapping randomly while following a higher level organization" is incomplete because it only encompasses completely stationary textures. It can even be argued that in reality no such texture can be observed: a synthesis algorithm strictly following this definition would thus be incomplete and of little use.

This means that a sound texture synthesis algorithm needs to be able to synthesize small indiscernible and random events but also singular, recognizable events, both harmonic (e.g. birds chirping) or not (e.g. crowd clapping). This extremely broad range of sounds is precisely what makes texture synthesis difficult and why common synthesis algorithms (for instance sinusoidal models) prove ill-suited for it, making it require dedicated ones.

Before presenting an overview of existing sound texture synthesis algorithm, we will first detail what we exactly expect from them.

In the case of textures, "re-synthesis" is a term as fitting as "synthesis": starting from an existing texture, the goal is usually to create a sound that is different from the original while still being identifiable as the same kind of texture, as if it had been recorded only moments later. Although this is the prime goal of the algorithm, this obviously does not exclude the possibility of manipulating the synthesized texture. For instance, it could be desirable to allow the algorithm to synthesize texture lasting any arbitrary length of time, or to be able to have the synthesis evolve throughout time, altering its properties or turning into another texture.

To achieve such a result, a broad variety of methods have been developed: for the needs of this article, we will split them into 3 different categories.

The first of those is \textit{physics-based synthesis}. In this one the goal is to first emulate the phenomenon at the source of the texture (say, the impact of a drop of rain on a flat surface) via a physically informed modelization of it. From there, one can simulate any number of events, dimensioning and randomizing it so as to fit the target texture, thus recreating a convincing physical simulation of the texture (see for instance \cite{o2002synthesizing}). While this method has the potential of being extremely controllable and allowing the manipulation of synthesis parameters that make physical sense, it also has the obvious flaw of not being flexible at all. Each algorithm will correspond to one and only one kind of texture: indeed, a physical model of the rain will prove poorly suited to synthesize a flock of birds twittering.

Then comes \textit{granular synthesis}. The original texture is first chopped into milliseconds-long audio particles, then reordered and concatenated to reconstruct a new texture (see for instance \cite{schwarz2007corpus}). While being quite versatile in the range of texture it is capable of re-synthesizing, this method is also heavily dependent on the choice of atom size and requires more complex reordering methods when one tries to synthesize a broader array of textures. In particular, reconstructing any foreground event lasting more than an atom will prove quite hard.

The last of those sound texture synthesis methods, to which our algorithm belongs, is \textit{parametric synthesis}. Here the general goal is to establish a set of parameters to describe textures with. If those parameters are well chosen, any two textures which parameters have the same values should sound like the same kind of texture without needing to be identical. Hence, one would only need to modify any sound until its parameters reach the values of a target texture's to re-synthesize it. While in theory this method is able to synthesize any and all kind of texture, in practice the quality and flexibility of the synthesis entirely depend on the choice of parameters. Indeed they need to describe the texture so that they hold enough information to re-synthesize a similar one, while not holding too much else the only way of creating a texture having the same parameter values will be by creating an exact copy. In addition to this, the parameters must be adapted to the widest possible range of textures.

This paradigm is notably used by McDermott and Simoncelli in \cite{mcdermott2011}, where a set of statistics extracted from the critical and modulation bands was used as parameters. This algorithm gave convincing results for a broad range of textures, but ran into trouble when trying to synthesize textures containing singular events. Inspired by the work of Gatys et al. \cite{gatys2015texture} who used the cross-correlations between the feature maps of a trained 2D convolutional neural network (CNN) as parameters to synthesize visual textures, several attempts to convert this approach to audio were made. In \cite{ulyanov2016}, Ulyaninov and Lebedev use the same principle applied to spectrograms, with the frequency dimension acting as color depth and using a 1D CNN, to synthesize textures with moderate success. In \cite{antognini2018synthesizing}, Antognini et al. add to this approach several constraints aimed at recreating rhythm with a better fidelity and increasing the diversity of results. While giving convincing results, this method also requires fine tuning in order to balance those constraints. Since changing target texture implies tuning those synthesis parameters, this makes the algorithm lose in flexibility. In addition to this, both this method and Ulyaninov and Lebedev's eventually output a spectrogram: it is then necessary to recover its phase and invert it to retrieve an audible time signal, using methods such as the Griffin and Lim algorithm \cite{griffin1984signal}. This phase recovery step is an added burden to the synthesis as it tends to downgrade the quality of the audio signal, even more so when working with complex sounds such as textures.

In this work, we present a new parametric texture synthesis based on the method of Gatys et al.\cite{gatys2015texture} which avoids the troubles of requiring any fine tuning or of going through spectrogram inversion while still working with a wide array of textures, including those presenting strong singular events. We then present a few examples to demonstrate its possibilities and proceed to discuss those results.

\section{Method}
\label{sec:method}

Following the principle of parametric synthesis, we first need to choose a set of parameters to represent a sound texture with.

\subsection{Parametrization}
\label{ssec:param}

\subsubsection{Pre-processing}
\label{sssec:preproc}

As our method is adapted from the work of Gatys et al. \cite{gatys2015texture} on 2D images, we require a 2D representation of our sound signal.

To this effect we work with log\hyph{}spectrograms. The log\hyph{}spectrogram $S$ is computed using the spectrogram $X$, taken as the magnitude of the short-term Fourier Transform (STFT) of the sound signal:

\begin{equation}
	S = \frac{\log (1 + C \times X)}{\log (1+C)}
\end{equation}

With $C$ a factor controlling compression: the larger $C$ is, the more details we will get at low amplitudes. This choice of normalization is made to ensure that the spectrogram is both compressed by the log function and comprised between 0 and 1.

For the rest of this article, any time-frequency matrix will have frequency as first axis, and time as second. For instance, $S(f,t)$ denotes the value of $S$ at the $f$-th frequency bin and $t$-th time sample.

\subsubsection{Network choice}
\label{sssec:net}

Seeing as Gatys et al. used a network trained for image recognition, our first intuition was to use an equivalent network for working on our spectrograms. As such we initially trained a simple deep 2D CNN on recordings taken from \href{https://freesound.org/}{freesound.org} for scene recognition in order to use it for synthesis. But in \cite{ustyuzhaninov2016texture}, Ustyuzhaninov et al. show that visual textures of the same quality as those obtained by Gatys et al. can be synthesized using a single-layer untrained CNN with various filter size instead of a trained CNN. This proves to still hold for sound textures: the network we use for the synthesized textures presented in this article is a single-layer untrained 2D CNN using filters of different size and ReLU activation. Its architecture is detailed in section \ref{ssec:netarch}. For generalization's sake, the rest of the method is nonetheless presented with a network that has $K$ layers (with $K$ being potentially more than 1), but is still valid when using a single-layer network.

It is worth noting that unlike the methods presented in \cite{ulyanov2016, antognini2018synthesizing}, we use the log\hyph{}spectrograms as 2D images with time and frequency replacing the two space dimensions and not as a 1D signal with frequency as depth, hence the need for 2D convolutions.

\subsubsection{Parameters computation}
\label{sssec:paramcomp}

Let us denote $F^{k}_{i, (x,y)}$ the value of the $i$-th feature map of the $k$-th layer at the position $(x,y)$. In \cite{gatys2015texture}, Gatys et al. use the Gram matrices of each layer of the network as parameters. The $(i,j)$ element of the gram matrix $G^k$ of $k$-th Gram marix is defined as the cross-correlation between the $i$-th and $j$-th feature maps of the layer:

\begin{equation}
	G^k (i,j) = \sum_{(x,y)} F^{k}_{i}(x,y) \times F^{k}_{j}(x,y)
\end{equation}

The parameter set is chosen as the list of gram matrices from $G^1$ to $G^K$. Although this proves a fine choice for visual textures, such parameters cannot be directly used in the case of sound textures. Indeed, those Gram matrices average all spatial information when performing a sum over all positions $(x,y)$, thus implying that the parametrization should be invariant in both direction. This does no translate well to sound, seeing as sound textures behave differently regarding time and regarding frequency: while we wish for a pseudo-stationarity over the time dimension, there is no reason for there to be any invariance to pitch-shifting. As such, we instead use the 3-dimensional tensors $H^k$ defined as:

\begin{equation}
	H^k (i,j, x) = \sum_{y} F^{k}_{i}(x,y) \times F^{k}_{j}(x,y)
\end{equation}

\begin{figure*}[ht]
\center
\includegraphics[width=5.5in]{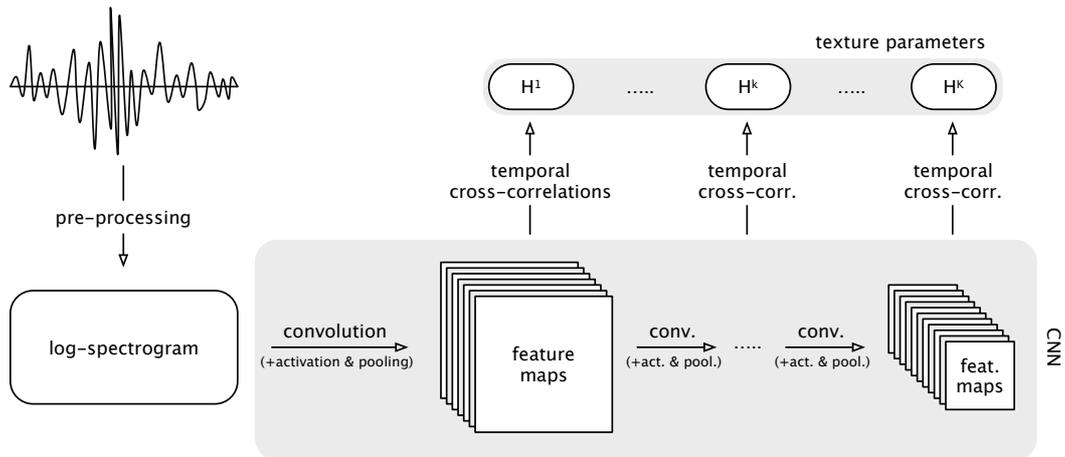}
\caption{\label{fig:gram}{\it Computation of the 3-dimensional parameter tensors from the temporal cross-correlations of the feature maps of a CNN, .}}
\end{figure*}

Defined this way, the tensors $H^k$ with $k\in[1,K]$ that we will use as parameters do indeed average all information from the time dimension, but keep the information regarding the frequency dimension intact, achieving our goal. The parameter extraction process is represented in figure \ref{fig:gram}.

\subsection{Texture loss}
\label{ssec:loss}

As mentioned in section \ref{sec:intro}, the main goal of parametric synthesis is to create a sound which has the same parameters values as a those of a target sound. Seeing as we now have a parameter set, we only need to define a quantitative error function which will then be minimized throughout the synthesis process. To that effect, we use a simple distance function between the two sets similarly to Ustyuzhaninov et al. \cite{ustyuzhaninov2016texture}:

\begin{equation}
	\mathcal{L} = \sum_k \frac{\Vert \tilde{H}^k - H^k \Vert_2}{\Vert \tilde{H}^k \Vert_2}
\end{equation}

With $\Vert . \Vert_2$ denoting the L2 norm, and the tilde denoting the target texture parameters.

\subsection{Optimization}
\label{ssec:opt}

The last step of the synthesis process is to create a sound signal which minimizes the texture loss. Since the chain of operations leading to the computation of the texture loss is differentiable (despite passing through the complex domain due to the STFT: see \cite{caracalla2017gradient} for further insight), we may use any optimization algorithm requiring the gradient of the error function to iteratively modify a sound until it reaches a satisfying minimum of the loss function.

\begin{figure*}[ht]
\center
\includegraphics[width=5.5in]{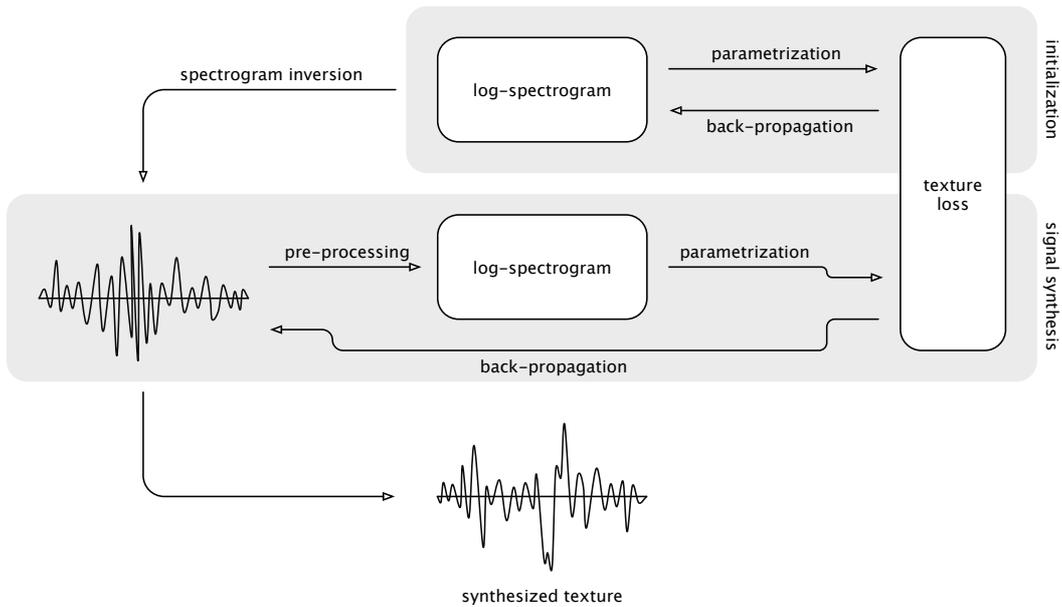}
\caption{\label{fig:synthesis}{\it Organization of the texture synthesis algorithm: a quick optimization on a log\hyph{}spectrogram is performed in the frequency-time domain and its result is inverted. It then serves as initialization for the main optimization performed on a sound signal.}}
\end{figure*}

We observed that performing the optimization on the log\hyph{}spectrogram is iteration-wise faster than performing the optimization directly on the sound signal (both of them being initialized using white noises). The texture loss appears to be easier to minimize when working in the time-frequency domain than when working directly in the time domain. To take advantage of that fact, we first perform a quick synthesis of a log\hyph{}spectrogram and invert it using a random phase matrix (which would correspond to performing one step of the Griffin-Lim algorithm): while this inversion raises the value of the texture loss, it still makes for a good initialization of the optimization in the time domain. This allows us to skip a major part of the optimization process on the sound signal. Once performed this optimization results in a sound signal which minimizes the texture loss, meaning its parameters values are close to those of the target texture. The whole synthesis process is illustrated in figure \ref{fig:synthesis}.

\subsection{Extension}
\label{ssec:ext}

\begin{figure*}[ht]
\center
\includegraphics[width=5.5in]{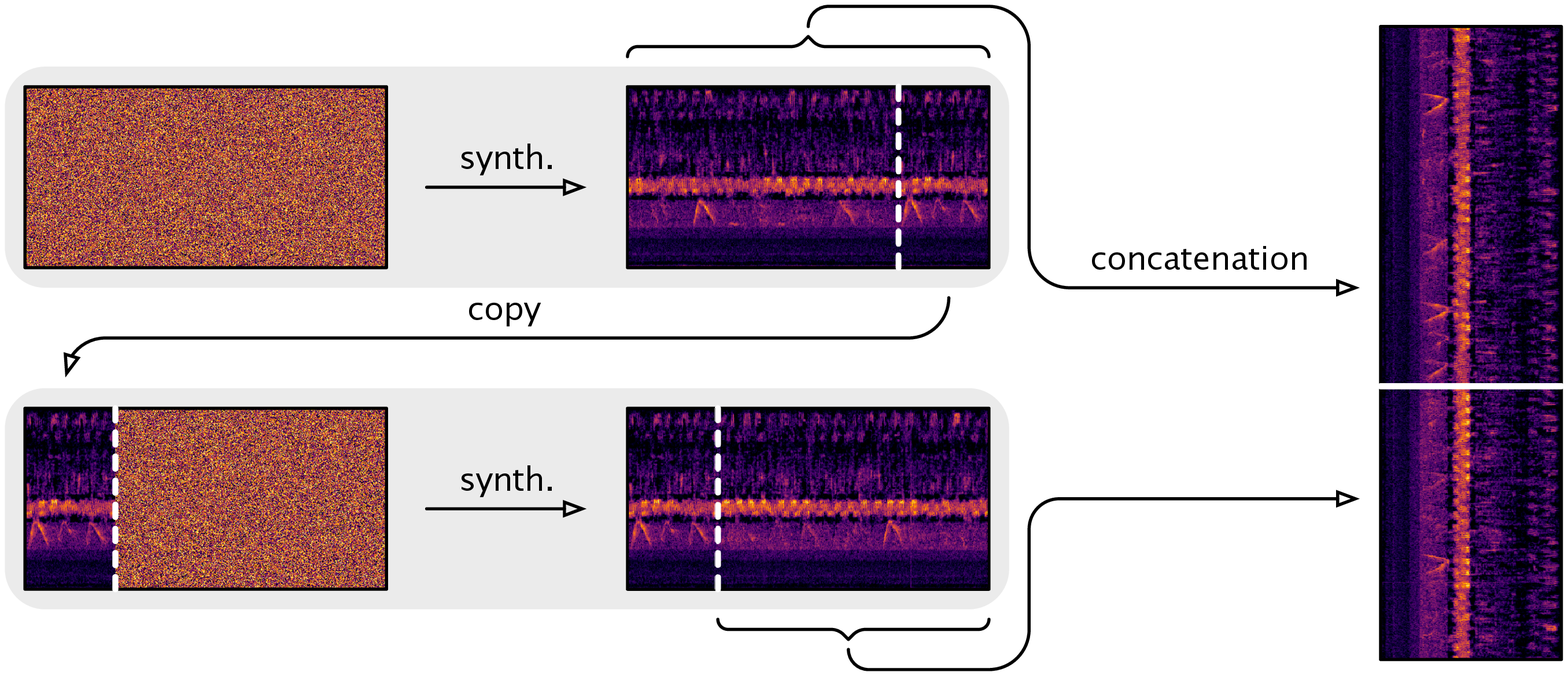}
\caption{\label{fig:exquisite}{\it Extension of a first synthesis by copying its end onto the start of a white noise signal. This signal is then used as initialization of another synthesis, while preventing the common part from being modified. The tiles are then concatenated to form a longer texture. Sound signals are represented by their log\hyph{}spectrogram for explanation's sake.}}
\end{figure*}

Once the basis for the texture synthesis has been set it becomes possible to develop on it, for instance by creating an indefinitely long sound texture. In order to do so, we use a principle resembling the "exquisite cadaver" game where one has to continue the drawing of another while only being seeing the border of the other's drawing. In our case we first synthesize an initial sound texture from a given target and copy the end of it onto the start of a white noise signal: we then perform another synthesis using this signal as initialization and the same target, while preventing the optimization to be performed on the section that was copied from the previous synthesis. This results in a continuous texture seamlessly extending on the copied part, thus being able to perfectly follow where the previous synthesis left off. We only need to concatenate the newly generated texture to the previous one to create a longer sound texture. This process can obviously be repeated any number of times so as to obtain a texture of any desired length. One iteration of this process is shown on figure \ref{fig:exquisite}, while an example of such a synthesis is available for listening at \href{http://recherche.ircam.fr/anasyn/caracalla/dafx19/extended_synthesis/}{recherche.ircam.fr\slash{}anasyn/caracalla\slash{}dafx19/extended\_synthesis/}.

\section{Results}
\label{sec:results}

\subsection{Synthesis parameters}
\label{ssec:synthpar}

The results presented in this section have been obtained using time signals sampled at $22050$ Hz, and a STFT with a window length of $512$ and hop size of $256$. The compression factor $C$ is set to $1000$. The sound signals all have a length of $262,400$ samples so that their log\hyph{}spectrograms are $1024$ frames long.

\subsection{Network architecture}
\label{ssec:netarch}

As mentioned in section \ref{sssec:net}, Ustyuzhaninov et al. \cite{ustyuzhaninov2016texture} show that using a single-layered untrained CNN yields equivalent results to using a trained CNN when synthesizing visual textures, and works best when the random CNN uses filters of several sizes. Taking inspiration from this, and after comparing results with a CNN trained for scene recognition we chose to use an untrained network made of a single multi-size convolutionnal layer. This layer is made of 128 square filters of each of the sizes [3, 5, 7, 11, 15, 19, 23, 27] with a stride of (1, 1) and zero-padding so that the differently-sized convolutions can then be stacked, followed by a ReLU activation layer.

\subsection{Optimization parameters}
\label{ssec:optpar}

The optimization algorithm chosen in this article is the L-BFGS algorithm. Starting from a white noise image, we perform 1000 iteration of it in the time-frequency domain to create the initialization of the time domain optimization. Said optimization is performed over 10000 iterations. Using a GeForce GTX 1080 Ti GPU, the whole process takes around an hour.

\subsection{Experimental results}
\label{ssec:respres}

\begin{figure*}[ht]
\center
\includegraphics[width=6.5in]{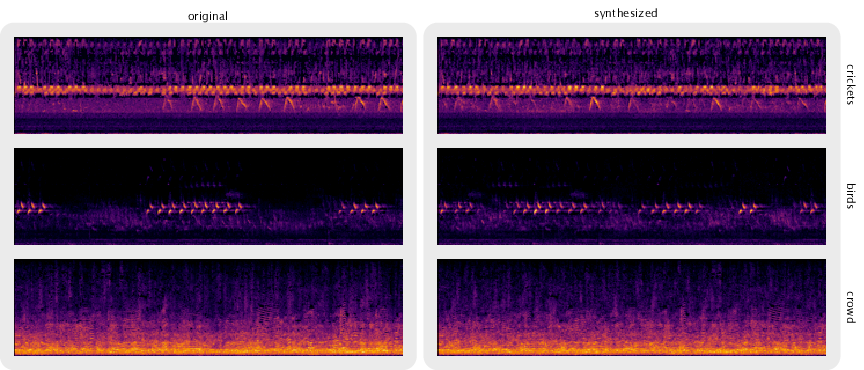}
\caption{\label{fig:results}{\it Log\hyph{}spectrograms of both originals textures (on the left) and synthesized textures (on the right) using the method presented in this article.}}
\end{figure*}

The log\hyph{}spectrograms of both target and synthesized textures are shown in figure \ref{fig:results} for three sounds : a wildlife scene with crickets chirping in the background and a bird singing in the foreground (recognizable to its inverted "v"-shaped patterns), birds singing both in the background and in the foreground (with one strongly standing out the mid-frequency range), and the hubbub of a crowd chatting. The audio signals of all three are available for listening at \href{http://recherche.ircam.fr/anasyn/caracalla/dafx19/synthesis/}{recherche.ircam.fr\slash{}anasyn/caracalla\slash{}dafx19/synthesis/}, along with other textures such as wind, bees and fire sounds.

\section{Discussion}
\label{sec:discussion}

\subsection{Results analysis}
\label{ssec:resana}

The strength of this texture synthesis algorithm lies in the fact that it manages to both synthesize background and foreground events convincingly. The crowd chatter, which includes no singular events, is as well recreated as the bird singing loudly in front of a flock.

Another advantage to it is the absence of any parameter tuning: unlike the method presented by Antognini et al. \cite{antognini2018synthesizing} where three losses need to be balanced through the optimization process and potentially from one texture to another, the texture loss used here is straightforward and requires no tuning. This way the algorithm can effectively be used without needing to take into account which texture is being synthesized.

In addition to this, because the final optimization is performed on the time signal our algorithm does not end with a spectrogram inversion (the harm it can bring is easily noticeable when trying to invert spectrograms of existing textures).

\subsection{Untrained vs. trained network}
\label{ssec:untrain}

As mentioned in section \ref{sssec:net}, both trained and untrained network have been used in visual texture synthesis with success. The main argument in \cite{gatys2015texture} is that the CNN trained for image recognition has learned filters adapted to common shapes encountered in images. This, coupled with the depth of the network, is what supposedly allows the network to recreate a large array of shapes when trying to reproduce the cross-correlations between the activations of its filters once the network has been fed an image (in this case, a visual texture). This argumentation is challenged by Ustyuzhaninov et al. in \cite{ustyuzhaninov2016texture}, who demonstrate that a single-layered untrained CNN can performed as well as a trained network given enough filters. This would tend to imply that given enough random filters, the space of shapes recreated when synchronizing some of those filters is wide enough to compensate for the lack of training.

This translates seamlessly to sound textures: while we first worked with networks trained for sound recognition, experiments with untrained network showed that they performed just as well. This being said, it could be interesting to explore the difference between the use of the two further: for instance, trained CNN might require less filters than untrained ones, thus making our parameter tensors lighter and the computations faster. The depth of the trained CNN might also help it capture correlations across distant events in the spectrogram. This would indeed be useful, seeing as the synthesized birds texture from figure \ref{fig:synthesis} shows that while the algorithm manages to reproduce the local pattern of bird cries well, it fails to reproduce the larger of pattern of groups of cries separated by gaps of a few seconds. This is also quite audible when listening to the attempt at (non-texture) singing voice synthesis: since the human voice is rich in harmonics, it spans over a large portion of the frequency spectrum. Because our algorithm does not enforce long-distance correlations, the upper harmonics are not synchronized with the lower ones, thus creating another high-pitched voice speaking on its own. In the fire synthesis, this is also clearly noticeable when looking at impacts: since those short and sharp events span over most frequencies, the algorithm has trouble generating them and mostly manages to recreate impacts that only span over part of the frequency axis, resulting in less convincing synthesized textures. This could potentially be solved even when using untrained network by choosing bigger filters, which would then "see" larger chunks of the log\hyph{}spectrogram at once.

\subsection{Border effect}
\label{ssec:bordereff}

Our texture synthesis presents one intriguing property: at the start and end of the sound, the synthesized texture is identical to the original one (for instance, this is slightly visible at the start of the log\hyph{}spectrogram of the birds texture from figure \ref{fig:synthesis}). In all of our synthesis, the leftmost and rightmost frames present exactly the same patterns in both original and synthesized textures. Although this effect is interesting, it dissipates quickly and only affects the time dimension: this means that even if we were to not get rid of this artifact, slightly cropping the start and end of the synthesized texture does completely discard it.

Gatys et al. \cite{gatys2015texture} noticed a similar effect in his visual texture synthesis where a distinctive part of the image was always reproduced around the same spot, and suggested that this effect originated from the zero-padding used in the convolution layers. To test this theory we experimented synthesizing textures using only "valid" convolutions (i.e. without any padding): the results still presented the same border artifacts, which would indicate that they do not originate from the zero-padding. We do not have any alternative explanation to present at the moment.

It is worth noticing that this effect lasts around the length of the biggest filter used (in our case, 27 frames): for now, this means that we need to choose a filter size large enough to ensure the good reproduction of correlations between events, while small enough so that the border effect doesn't spoil too much of the interior of the synthesized texture.

\subsection{Computation time}
\label{ssec:comput}

As mentioned in section \ref{sssec:paramcomp}, computation time is for now far behind real-time since it takes around an hour to synthesize 12 seconds of audio with one GPU. While tedious for now, this process could potentially be alleviated by removing as much redundant information from the target parameters as possible (as of now, we have around $\mathtt{\sim}100$M parameters in the parameter tensors $H_k$).

Using the same network as described in section \ref{ssec:netarch}, we tried removing all cross-correlations between filters of different size from the parameter tensor: this dropped the number of parameters to $\mathtt{\sim}16$M without altering the quality of synthesized textures. Another lead would be to use a trained CNN instead of an untrained one, seeing as trained filters should prove efficient at describing patterns without needing to be as numerous as in an untrained CNN. We believe it should also be possible to drop the number of parameters even lower, for instance by using principal component analysis to select which cross-correlations need to be imposed over the synthesized signal as Gatys et al. \cite{gatys2015texture} did. 

Another potential lightening of the algorithm could come from changing signal representation: so far we have used log\hyph{}spectrograms, but using another time-frequency representation could be greatly beneficial. For instance, using mel bands instead of the raw frequency bins of the spectrogram would reduce the number of parameters while staying perceptually sensible.

\section{Conclusions}

We introduced a new parametric texture synthesis based on the work of Gatys et al. \cite{gatys2015texture} in visual textures: using the temporal cross-correlations between the feature maps of a CNN as parameters, we iteratively modify a sound signal until the values of its parameters reach those of a target texture. While the input of the CNN is the log\hyph{}spectrogram of the sound, the optimization process is made directly in the time domain so as to avoid any phase recovery step in the synthesis.

The algorithm yields convincing results on a wide array of texture, even if they include singular events in the foreground. It can be straightforwardly applied without requiring the fine tuning of synthesis parameters from one texture to another. Its major flaws for now lie in its long computation time and its trouble re-synthesizing correlations of events far apart in the log\hyph{}spectrogram. A number of possible ways to address the first issue have been presented, for instance by subsampling the parameters tensor and altering the time-frequency representation. As for the second, the influence of the CNN architecture, and most notably the shape of its filters, are currently being investigated.

\nocite{*}
\bibliographystyle{IEEEbib}
\bibliography{DAFx19_tmpl} 

\end{document}